\documentclass[twocolumn,showpacs,preprintnumbers,amsmath,amssymb,floatfix]{revtex4}
\usepackage{graphicx}
\usepackage{dcolumn}
\usepackage{bm}
\usepackage{float}

\newcommand{\etal}{\mbox{\em et al.}}
\newcommand{\degree}{\mbox{$^\circ$}}
\newcommand{\gevc}{\mbox{GeV/c}}
\newcommand{\mevc}{\mbox{MeV/c}}
\newcommand{\gevctwo}{\mbox{(GeV/c)}^2}
\newcommand{\eep}{\mbox{(e,e$^\prime$p)}}

\newcommand{\deep}{\mbox{D(e,e$^\prime$p)n}}

\newcommand{\qvec}{\mbox{$\vec q$}}
\newcommand{\qlab}{\mbox{$q_{\mathrm{lab}}$}}
\newcommand{\ppi}{\mbox{$p_i$}}

\newcommand{\ppf}{\mbox{$p_f$}}

\newcommand{\ppm}{\mbox{$p_{m}$}}

\newcommand{\Qtwo}{Q^2}
\newcommand{\dif}{\mbox{d}}
\newcommand{\thetapq}{\mbox{$\theta_{pq}$}}

\newcommand{\thetae}{\mbox{$\theta_e$}}

\newcommand{\thetanpcm}{\mbox{$\theta_{np}^{\mathrm{c.m.}}$}}

\newcommand{\vqcm}{\mbox{$q_{\mathrm{c.m.}}$}}
\newcommand{\dfsig}{\mbox{$\frac{\dif^5\sigma}{\dif\omega\dif\Omega_e\dif\Omega_{p}}$} }

\newcommand{\sigep}{\mbox{$\sigma_{ep}$}}

\newcommand{\rlt}{\mbox{$R_{LT}$}}

\newcommand{\fl}{\mbox{$f_{L}$}}
\newcommand{\ft}{\mbox{$f_{T}$}}
\newcommand{\flt}{\mbox{$f_{LT}$}}
\newcommand{\ftt}{\mbox{$f_{TT}$}}

\newcommand{\rol}{\mbox{$\rho_{LT}$}}

\begin{document}

\title{The $f_{LT}$ Response Function of D(e,e'p)n at  $Q^2=0.33$ $\gevctwo$}

\author{W.U.~Boeglin} \altaffiliation{Present Address: Florida
  International University, Miami, USA} 
\author{H.~Arenh\"ovel}
\author{K.I.~Blomqvist}
\altaffiliation{Present address: DANFYSIK, Jyllinge, Denmark}

\author{R.~B\"ohm} \author{M.~Distler} \author{R.~Edelhoff} 
\author{I.~Ewald} 
\altaffiliation{Present address: Renewable Energies,
  Koenig-Konrad-Strasse 2, 55127 Mainz, Germany}

\author{R. Florizone} 
\altaffiliation{Present address: Canadian Light Source Inc.,
  University of Saskatchewan, 101 Perimeter Road, Saskatoon, SK.,
  Canada. S7N 0X4} \author{J.~Friedrich} \author{R.~Geiges}

\author{M.~Kahrau}
\author{M.~Korn} 
\author{H.~Kramer} 
\altaffiliation{Present address: TLC GmbH, Wiesbaden, Germany}

\author{K.W.~Krygier} \author{V.~Kunde} 
\altaffiliation{Present address: Mannesmann Autocom, D\"usseldorf,
  Germany} \author{A.~Liesenfeld} 
\author{K.~Merle} \author{R.~Neuhausen} \author{E.A.J.M.~Offermann}
\altaffiliation{Present address: Renaissance Technologies, Stony
  Brook, USA} \author{Th.~Pospischil} \author{A.W.~Richter}
\altaffiliation{Present address: TLC GmbH, Wiesbaden, Germany}
\author{G.~Rosner} \altaffiliation{Present address: Department of
  Physics and Astonomy, University of Glasgow, Glasgow, G128QQ,
  Scotland, IK} \author{P.~Sauer} \altaffiliation{Present address: TLC
  GmbH, Wiesbaden, Germany} \author{S.~Schardt} 
\author{A. Serdarevic} \altaffiliation{Present address: 126 Cliff Rd,
  Port Jefferson, NY, USA} \author{A.~Wagner} 
\author{Th.~Walcher} \author{S.~Wolf} 
\affiliation{Institut f\"ur Kernphysik,\\
  Johannes  Gutenberg-Universit\"at,\\
  J.J.-Becher-Weg 45,\\
  D-55099 Mainz, Germany} \author{J.~Jourdan} \author{I.~Sick}
\affiliation{Dept. f\"ur Physik und Astronomie, \\
  Universit\"at Basel, \\
  CH4056 Basel, Switzerland} \author{M.~Kuss} 
\altaffiliation{Present address: INFN Pisa, Italy}
\affiliation{Institut f\"ur Kernphysik, \\
  TH Darmstadt, \\
  D-64289 Darmstadt, Germany} \author{M.~Potokar}
\author{A.~Rokavec} 
\author{B. Vodenik} 
\author{S.~Sirca}
\affiliation{Institute Jo\v{z}ef Stefan, \\
  University of Ljubljana, \\
  SI-61111 Ljubljana, Slovenia}

\date{\today}

 \begin{abstract}
   The interference response function $\flt$ ($\rlt$) of the $\deep$
   reaction has been determined at squared four-momentum transfer 
   $\Qtwo = 0.33$ $\gevctwo$ and for missing momenta up to $\ppm= 0.29$
   $\gevc$. The results have been compared to calculations that reproduce
   $\flt$ quite well but overestimate the cross sections by 10 - 20~\% for
   missing momenta between 0.1 $\gevc$ and 0.2 $\gevc$.
\end{abstract}


\maketitle

%
%
\section{Introduction}
The deuteron is an ideal system to investigate fundamental problems in
nuclear physics such as the ground state and continuum wave functions
and the structure of the electromagnetic current operator. In
addition, interaction effects such as meson exchange currents (MEC),
and isobar configurations (IC) can be studied.

The deuteron structure can be calculated with very high 
accuracy therefore providing a testing ground for various 
models of the nucleon-nucleon force and subnuclear degrees of freedom. 

The exclusive deuteron electro-disintegration cross section has been
measured at several laboratories during the past 25
years (References to these experiments can be found in the text
below.). However there are only a few experiments where the individual
response functions have been separated. 
In this paper we report a measurement of the $\flt$ ($\rlt$) response
function, extending the kinematical area where experimental data are
available.

The $\deep$ reaction can be most easily interpreted within the framework
of the plane wave impulse approximation (PWIA). In this approximation
the cross section is written  as follows:
\begin{eqnarray}
\dfsig  =  \kappa \cdot \sigep \cdot S(\ppi)\,.
\label{eq:sig_pwia}
\end{eqnarray}
Here, $\sigep$ describes the elementary electron proton (off-shell)
cross section for scattering an electron off a moving bound proton
\cite{deFor83}. The factor $\kappa$ is a kinematical factor, and
$S(\ppi)$ is the spectral function which describes the probability of
finding a proton with an initial momentum $\ppi$. In this
approximation, the initial momentum of the proton is opposite and
equal in magnitude to the missing momentum $\ppm$, the momentum of the
recoiling, non-observed neutron.

\begin{figure}
\includegraphics[width=0.5\textwidth,clip=true]{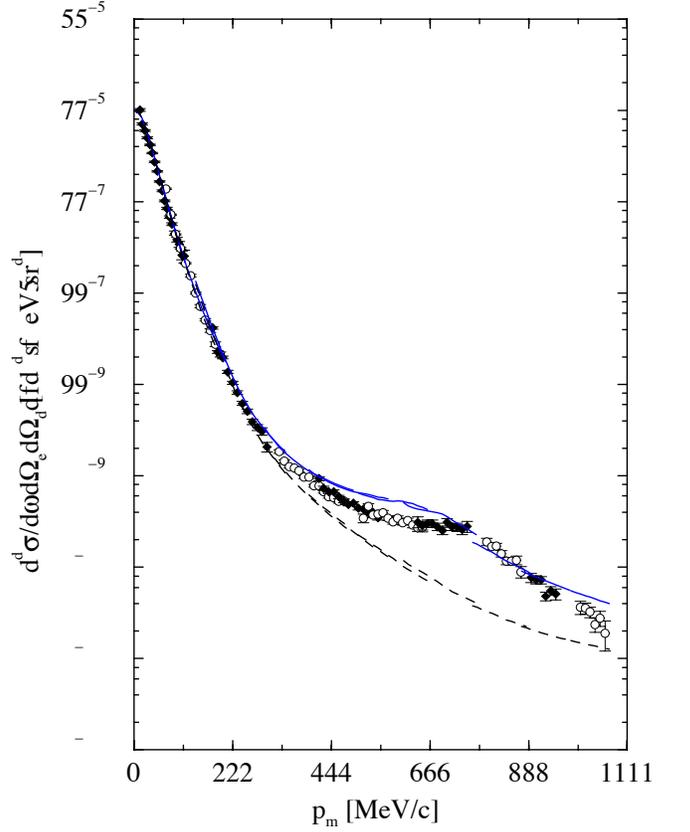}
\caption{\label{fig:sig_all} The experimental $\deep$ cross section as
  a function of missing momentum measured at MAMI for $\Qtwo = 0.33$
  $\gevctwo$\cite{bl98} compared to calculations~\cite{ArL05} 
  with (solid curve) and without (dashed curve) MEC and IC. Both
  calculations include FSI. The low $\ppm$ data have been re-analyzed
  and used in this work to determine $\flt$ (color online).}
\end{figure}

Several experiments explored the $\deep$ cross section over a wide
range of missing momenta at small to medium momentum transfers
\cite{Bern81,Tur84,bl98,Ulm02}. The focus of these measurements was
the exploration of the momentum distribution within the plane wave
impulse approximation (for a theoretical analysis of the Saclay
experiment~\cite{Bern81} see~\cite{Are82}). 
It has been found, however, that with
increasing recoil momentum FSI and, related to the corresponding
energy transfer, MEC and IC contributions increase
dramatically. Figure~\ref{fig:sig_all} shows the $\deep$ cross section
measured at MAMI~\cite{bl98} together with a 
calculation that include FSI, MEC, and IC~\cite{ArL05}. One can see
that the cross section is well reproduced up to $\ppm = 350$ $\mevc$
by a calculation that includes FSI. At higher {$\ppm$} there are
significant discrepancies between experiment and the FSI
calculation. If additionally MEC and IC are included the agreement
improves considerably but significant discrepancies remain. The
largest deviations occur at energy transfers where large virtual
delta excitation contributions are expected.

When these additional contributions are taken into account the cross
section cannot be factorized in this simple way anymore and one has to
use the full one photon exchange approximation. Within this limit, the $\eep$
cross section can be written as follows:
\begin{align}
\frac{\dif^5\sigma}{\dif\omega^{\mathrm{lab}}\dif\Omega_e^{\mathrm{lab}}
\dif\Omega_{p}^{\mathrm{lab}}} = \sigma_{Mott}( &v_LR_L + v_TR_T \nonumber \\
                    +&v_{LT}R_{LT}\cos{\phi} \nonumber \\
                    +&v_{TT}R_{TT}\cos{2\phi})\,.
\label{eq:geneep}
\end{align}
The functions $R_{x}$ ($x\in\{L,T,LT,TT\}$) are re\-sponse functions. They
consist of combinations of transition matrix elements of the
components of the electromagnetic current operator and contain the
structure information; the incident and the scattered electrons
are described as plane waves. The angle $\phi$ is the angle between the
electron scattering plane and the reaction plane, defined by the
momentum of the ejected nucleon and the momentum transfer.

In view of the fact that the theoretical calculation~\cite{ArL05} is based on
an evaluation of the responses in the final $np$-c.m.\ system using the
following form of the differential cross section (note that
$\phi=\phi_{np}^{\mathrm{c.m.}})$         
\begin{align}
{\frac{\dif^5\sigma}{\dif\omega^{\mathrm{lab}}\dif\Omega_e^{\mathrm{lab}}
\dif\Omega_{np}^{c.m.}}}
 = C (&\rho_Lf_L(\theta_{np}^{\mathrm{c.m.}})+\rho_Tf_T(\theta_{np}^{\mathrm{c.m.}}) \nonumber \\
+&\rho_{LT}f_{LT}(\theta_{np}^{\mathrm{c.m.}})\cos\phi_{np}^{\mathrm{c.m.}} \nonumber \\
+&\rho_{TT}f_{TT}(\theta_{np}^{\mathrm{c.m.}})\cos2\phi_{np}^{\mathrm{c.m.}})\,, 
\label{eq:deutcro}
\end{align}
we now switch to the response functions $f_x$ with
$x\in\{L,T,LT,TT\}$. 
Using the relations
\begin{equation}
C=\frac{\eta}{6\pi^2\alpha Q^2}\,\sigma_{Mott}\,,\\
\end{equation}
where $\alpha$ denotes the fine structure constant and
$\eta=\tan^2(\theta_e/2)$, 
and 
\begin{alignat}{2}
\rho_L &= \tilde{\beta}^2\frac{Q^2}{2\eta} v_L\,,  &
\qquad \rho_T &= \frac{Q^2}{2\eta}v_T \,,\nonumber \\ 
\rho_{LT} &= \tilde{\beta}\frac{Q^2}{2\eta} v_{LT}\,, &
\qquad  \rho_{TT} &= \frac{Q^2}{2\eta}v_{TT} \,,
\label{eq:conv_v}
\end{alignat}
where $\tilde{\beta} = \frac{\qlab}{\vqcm}$ 
epresses the boost from the lab to the c.m.\ system,
one obtains the relations between the response
functions $R_x$ and the $f_x$ as follows:
\begin{alignat}{2}
\frac{\tilde{\beta}^2{\cal J}}{12 \pi^2\alpha}f_L &= R_L\,,&
\qquad \frac{\tilde{\beta}{\cal J}}{12 \pi^2\alpha}f_{LT} &=
R_{LT}\,,\nonumber \\ 
\frac{{\cal J}}{12 \pi^2\alpha}f_{T}& = R_{T}\,,&
\qquad \frac{{\cal J}}{12 \pi^2\alpha}f_{TT} &= R_{TT}\,,
\label{eq:conv_r} 
\end{alignat}
with ${\cal
  J}=|\partial\Omega^{\mathrm{c.m.}}_{np}/\partial\Omega^{\mathrm{lab}}_p|$ 
as Jacobian.

A full separation of all four response functions requires at
least one cross section to be measured with the proton detected out of
the electron scattering plane. This has been achieved at MIT--Bates using
 the Out--Of--Plane spectrometer (OOPS) system~\cite{Zho01}
 and at NIKHEF~\cite{Pel97} using the HADRON detectors. For an
 overview of results see~\cite{Zho98}.

Simpler in-plane measurements allow one to
separate, $\fl$, $\ft$ and $\flt$. The response function which is
easiest to determine is 
$\flt$ since in this case the electron momentum can remain constant, and one
only has to scan the proton momentum such that the $\eep$ cross sections can be
measured at $\phi = 0\degree$ and at $\phi = 180\degree$. 

In-plane separations 
have been carried out at recoil momenta between 0 and 220~MeV/c and at lower
$\Qtwo$ values at several laboratories and the published results can
be found in references~\cite{vdSch91,vdSch92,Jor96, Kas97}.  The momentum
transfer dependence of $\fl$, $\ft$ and $\flt$ has been measured at
Saclay for recoil momenta between 0 and 150~MeV/c \cite{Duc94}. At
SLAC, cross sections and $\flt$ have been determined at large momentum
transfers for recoil momenta up to 200 $\mevc$ \cite{Bul95}.

In this paper we report on the determination of $\flt$ close to the
quasi-free peak (for Bjorken variable $0.84 < x < 1$), at an average 
$\Qtwo$ of 0.33 $\gevctwo$ for missing momenta up to 290 $\mevc$.

\section{Experimental Details}

The experiment has been carried out at the three-spectrometer facility
\cite{A1} at the Mainz microtron MAMI using spectrometer B to
detect electrons and spectrometer A to measure protons. The incident
beam energy was $E_{inc} =$ 855.11~MeV, and the electron scattering
angle was kept constant at $\theta_e = 45$\degree. The momentum
acceptance of the electron spectrometer was $\Delta p/p = \pm 7.4~\%$
and the one of the proton spectrometer $\Delta p/p = -5, +15~\%$ with
respect to the corresponding reference momenta. The rectangular
entrance slit of the electron spectrometer defined an angular
acceptance of $\Delta \theta_e = \pm 20$~mr in the scattering
(horizontal) plane, and $\Delta \phi_e = \pm 70$~mr in the vertical
plane.  The proton spectrometer had an acceptance of $\Delta \theta_p
= \pm 75$~mr (horizontal) and $\Delta \phi_p = \pm 70$~mr
(vertical). The momenta of the outgoing protons varied between 531
MeV/c and 627 MeV/c. 

For the determination of $\flt$, three spectrometer
settings were selected for each central value of $\phi = 0\degree$ and
$\phi = 180\degree$ and one setting where the proton spectrometer was
centered around the direction of $\qvec$.
\begin{table} 
\caption{\label{tab:kin_cen} Central Spectrometer Settings} 
\begin{ruledtabular} 
\begin{tabular}{ccccc} 
kinematics & $\theta_e$ & $p_e$ & $\theta_p$ & $p_p$ \\ 
  &{\small $(\degree)$} & {\small $\rm (MeV/c) $} & {\small $(\degree)$} & {\small $\rm (MeV/c) $}\\ \hline 
 rlt00 & 45.00 & 657.1 & 49.85 & 608.4 \\ 
 rlt020 & 45.00 & 657.2 & 39.84 & 599.1 \\ 
 rlt040 & 45.00 & 657.6 & 29.68 & 569.5 \\ 
 rlt048 & 45.00 & 656.0 & 25.17 & 554.0 \\ 
 rlt1800 & 45.00 & 657.1 & 49.85 & 608.4 \\ 
 rlt18020 & 45.00 & 657.2 & 59.98 & 599.1 \\ 
 rlt18040 & 45.00 & 657.6 & 70.13 & 569.5 \\ 
 rlt18048 & 45.00 & 657.5 & 74.54 & 551.3 \\ 
\end{tabular} 
\end{ruledtabular} 
\end{table} 

\begin{figure}
\includegraphics[width=0.45\textwidth,clip=true]{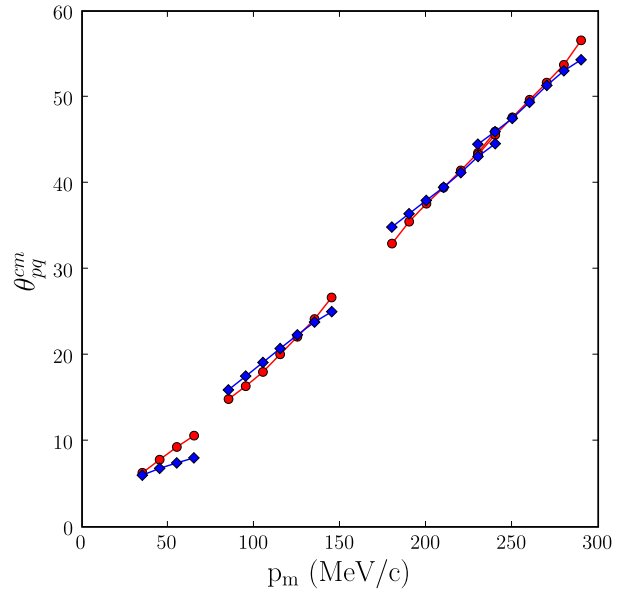}
\caption{\label{fig:thcm_pm} The relation between missing momentum and
the angle between the ejected proton and the momentum transfer in the
center of mass. The red circles correspond to the spectrometer
settings centered around $\phi = 0\degree$ and the blue diamonds
correspond to $\phi = 180\degree$ (color online).}
\end{figure}

 These proton spectrometer
settings corresponded to the center of mass angles
$\thetanpcm = 0\degree, 20\degree, 40\degree$ and $48\degree$. For a
given, fixed electron kinematics, a variation of $\thetanpcm$ also
corresponds to a change of the missing momentum. The relation between
$\thetanpcm$ and $\ppm$ for this experiment is shown in
figure~\ref{fig:thcm_pm}. The central settings of the spectrometers 
are listed in table~\ref{tab:kin_cen}.

The energy and momentum transfer, $\omega$ and $\qvec$ were kept
constant, centered at 200~MeV and at 600 MeV/c respectively. Both of
these quantities varied slightly by a few MeV/c due to the large
acceptances of the spectrometers.
The maximum value of
\thetanpcm was determined by the smallest angle with respect to the
beam, that the out-going protons could be detected for the given
electron kinematics. A detailed list of kinematics can be
found in the appendix in tables~\ref{tab:00_ks} -- \ref{tab:18048_ks}.

We used a liquid-deuterium target consisting of a cylindrical target
cell with a diameter of 2 cm made of HAVAR and a wall thickness of
$6.5~\mu$m or 10~mg/cm$^2$. The deuterium target thickness was
310~mg/cm$^2$.  The liquid deuterium was continuously circulated by
means of an immersed fan, thus preventing the liquid at the
intersection with the electron beam from boiling. Since the beam
diameter was typically of the order of 0.2 mm, the beam was rastered
horizontally by $\pm 3.5$ mm and vertically by $\pm 2.5$ mm with a
frequency of 3.5~kHz horizontally and 2.5 kHz vertically to further
reduce the risk of boiling.  The current in the raster coil was
measured on an event by event basis which allowed us to reconstruct
the beam position for each event in order to correct for energy losses
in the target. After applying the necessary kinematical corrections we
obtained at low missing momenta a missing energy resolution of
0.45~MeV (FWHM) which degraded to 2~MeV with increasing recoil
momentum, as one has to include increasingly large recoil energies in
the calculation of the missing energy.  With this target system, beam
currents between 2~$\mu$A and 40~$\mu$A could be used.

\begin{figure}
\includegraphics[width=0.5\textwidth]{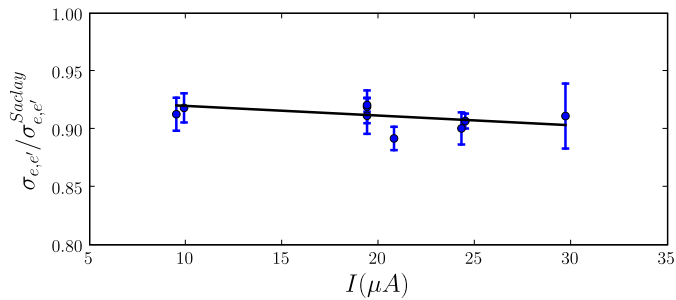}
\caption{\label{fig:norm} The ratio between the D(e,e')D cross
  section from this experiment and the interpolated elastic cross
  sections, measured at ALS in Saclay~\cite{Auf85,Pla90}. This ratio
  has been used to normalize all coincidence cross sections (color online).}
\end{figure}

The effective target thickness has been determined using elastic
scattering via D(e,e$'$D) measurements and normalizing to the data of
Platchkov et~al.~\cite{Pla90} and Auffret et~al.~\cite{Auf85}. These
normalization measurements were performed in regular intervals during
the experiment. From fitting a line to the ratio of the measured cross
section in this experiment to the Saclay data, we found a current
dependence of the target thickness of 0.1~\%/$\mu$A (figure~\ref{fig:norm}). 

The normalization factor varied between 1.082 at 5~$\mu$A and 1.12 at
40~$\mu$A. Contributions to this factor are the change of the effective
target thickness due to the horizontal rastering of the beam position
(4~\%), the 3~\% hydrogen admixture to the deuterium gas, and losses of
recoil deuterons due to nuclear reactions, estimated to be less than
3~\%.

The systematic error of the measured cross sections has been
determined to be about 6.2~\%.  It contains contributions from the
uncertainty in the elastic deuteron cross section (2~\%), estimated
deuteron losses (2.5~\%) and the uncertainty of the normalization
factor due to the statistical error in the D(e,e$'$D) cross section
(1.7~\%).

The error due to the uncertainties in the kinematic variables such as
beam energy, beam direction, electron momentum and direction and
proton direction was estimated for each bin. Their values lie between
0.3~\% and 4~\% depending on the kinematics. The largest errors are
found for the setting where the protons are almost parallel (central
setting where $\thetanpcm = 0\degree$) to the momentum transfer and the
$\eep$ cross section is dominated by $\fl$ and $\ft$. For this setting the
experimental cross sections were dominated by statistical errors and
the extracted $\flt$ provides just an upper limit. For the next
setting  ($\thetanpcm \geq 20\degree$) they
are of the order of 1~\% and below for the rest of the data. We have
added these errors in quadrature to the statistical ones.

\section{Determination of  $\flt$}
In order to extract the cross sections, the data have been binned in two
dimensions, missing energy and missing momentum. The spectra have
subsequently been radiatively unfolded and corrected for the coincidence phase
space acceptance. For each bin in missing momentum, we obtained the
cross section by integrating over missing energy, where the $\deep$
reaction produces a peak at 2.25 MeV.
\begin{figure}
\includegraphics[width=0.5\textwidth]{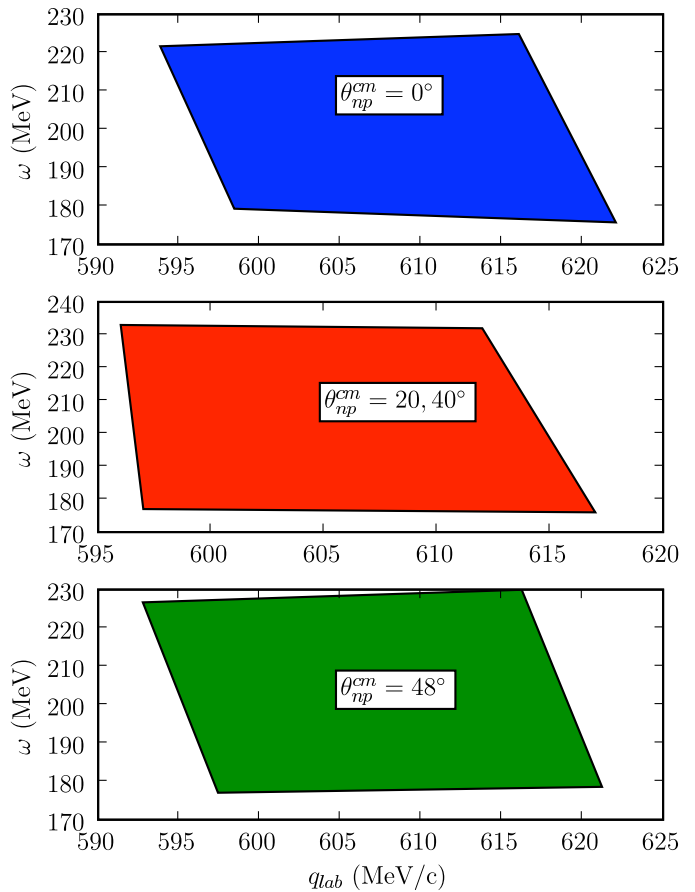}
\caption{\label{fig:cuts} Cuts in $\qlab$ and $\omega$ applied to the
  different kinematical settings. The same cuts have been applied to
  the $\phi = 0\degree$ as well as to the $\phi = 180\degree$ settings (color online). }
\end{figure}

\begin{table} 
\caption{\label{tab:cuts} Angle cuts applied at the kinematical settings} 
\begin{ruledtabular} 
\begin{tabular}{ccccc} 
kinematics & $\phi_{min}$& $ \phi_{max}$ & $\thetapq_{min}$ & $\thetapq_{max}$  \\ 
  &{\small $(\degree)$} & {\small $(\degree)$}  &{\small $(\degree)$} & {\small $(\degree)$} \\ \hline 
 rlt00    & -60 & 60  & 1 & 10 \\ 
 rlt020   & -35 & 35  & 5 & 15 \\ 
 rlt040   & -15 & 15  & 15 & 25 \\ 
 rlt048   & -15 & 15  & 20 & 30 \\ 
 rlt1800  & 120 & 240 & 1 & 10 \\ 
 rlt18020 & 145 & 215 & 5 & 15 \\ 
 rlt18040 & 165 & 195 & 15 & 25 \\ 
 rlt18048 & 165 & 195 & 20 & 30 \\ 
\end{tabular} 
\end{ruledtabular} 
\end{table} 

The large acceptances of the spectrometers lead to large regions in
the kinematic variables that have been sampled at each spectrometer
setting. Cuts in $\qvec, \omega, \thetapq$ (the angle between the
ejected proton and the momentum transfer), and $\phi$ have been
applied (figure~\ref{fig:cuts} and table~\ref{tab:cuts}) in order to
have well defined the kinematical regions sampled in each spectrometer
setting.  In spite of these cuts different missing energy and missing
momentum bins average the cross section over smaller, different
kinematic areas.

\begin{figure}
\includegraphics[width=0.485\textwidth,clip=true]{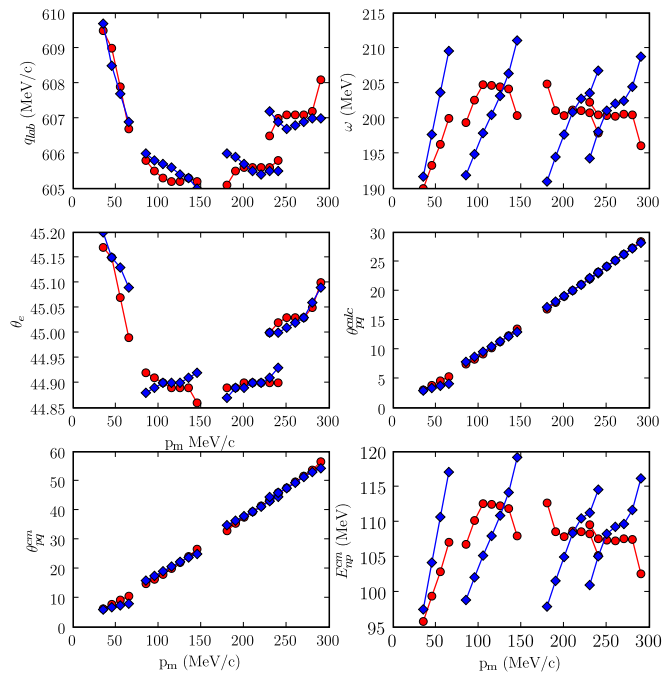}
\caption{\label{fig:kin_var}Averaged kinematic variables as a
  function of missing momentum. The red circles correspond to $\phi =
  0\degree$ and the blue diamonds correspond to $\phi = 180\degree$ (color online). }
\end{figure}

In order to take this into account we used a Monte-Carlo calculation,
including the full theoretical model, to determine the average of the
relevant kinematical variables for each missing energy/missing
momentum bin. The average variables evaluated this way were the
electron scattering angle ($\thetae$), the momentum and energy
transfers ($\omega_\mathrm{lab}, \qlab$) and the final proton momentum
$\ppf$. From 
these averaged quantities and the missing momentum we subsequently
calculated the average angle between the outgoing proton and the
momentum transfer $\thetapq$. This quantity could also have been
obtained directly from the Monte-Carlo calculation; however, the
$\deep$ kinematics would then have been over-determined and the
averaging process would lead to inconsistent kinematic results. We
therefore selected to calculate the average value of $\thetapq$ from
the averaged values for $\ppf$, $\qlab$ and $\ppm$.  The average
kinematic variables as a function of missing momentum are shown in
figure~\ref{fig:kin_var}.

From the same calculation we also obtained the averaged values for the
Mott cross section, the recoil factor, the kinematical factors for the
response functions (i.e.\ the density matrix for the virtual photon
polarization $\rho_{x}$, in eq.~(\ref{eq:deutcro})) and 
averages of $\cos{\phi}$ and
$\cos{2\phi}$. $\rho_{x}$ can also be calculated from the averaged
kinematical values associated with each data bin. 

From the general $\deep$ cross section in~(\ref{eq:deutcro})
one sees that the interference response functions can be extracted
from the $\phi$ dependence of the cross section. Most of the data
taken in this experiment are in or close to the electron scattering
plane, with $\phi$ angles distributed around $\phi=0\degree$ and
$\phi=180\degree$. We have decided to extract $f_{LT}$ from the 
cross section difference: 
\begin{equation}
  \sigma_{LT} = \frac{1} 
    {\overline{\cos{\phi_0}} - \overline{\cos{\phi_1}}}(\sigma_{0} -
  \sigma_{180})  
\label{eq:sig01exp}
\end{equation}
\begin{equation}
  \sigma_{LT} = C  \cdot \overline{\sigma_{Mott}}  \cdot
  \overline{\rol} \cdot\flt
\label{eq:sig01}
\end{equation}
Here $\overline{\cos(\phi_0)}$ and $\overline{\cos(\phi_1)}$ are
averages for the settings centered around $\phi = 0\degree$ and $\phi =
180\degree$ respectively, $C$ contains all normalization factors and
$\overline{\sigma_{Mott}}$ is the averaged Mott cross section for the bin
considered. This separation required that all kinematical
variables except $\phi$ are identical. This is in general not the case
as can be seen from figure~\ref{fig:kin_var}. Only the central bin
approximately satisfies this condition. 

In addition to the small overlap between the two $\phi$ settings, the
cross section determined for each missing momentum bin differs
slightly from the cross section corresponding to the averaged
kinematics associated with this bin. This
difference will also introduce a systematic error in the extracted
response function and needs to be corrected.

To correct for these effects we have calculated the averaged cross
section for each missing momentum bin. For one set of calculations we
used PWIA and the momentum distribution calculated with the Paris
potential, for the other set of calculations we interpolated the 
theoretical response functions that include FSI, MEC
and IC. These two calculations allow one also to estimate the model
dependence for correction factors derived for the effects above.

\begin{figure}
\includegraphics[width=0.5\textwidth,clip=true]{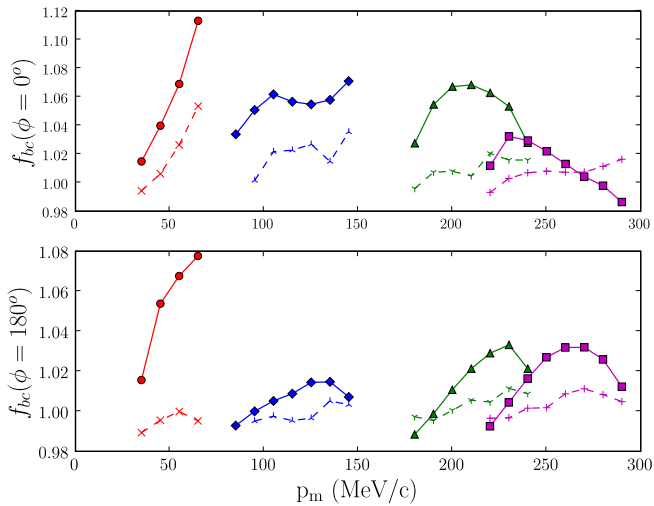}
\caption{\label{fig:bin_center_all} Ratio between the averaged
  $\deep$ cross section for each missing momentum bin and the cross
  section calculated using the corresponding averaged kinematics. Top:
  ratios for $\phi = 0\degree$. Bottom: ratios for $\phi =
  180\degree$. The full lines joining the solid symbols are calculated
  using the full theory while the dashed lines are
  calculated using PWIA. The ratios $f_{bc}$ calculated with the full
  calculation are used to correct the experimental cross sections as
  described in the text (color online).}
\end{figure}

The ratio between the cross section calculated for the averaged
kinematics ($\sigma^{calc}_{kin_{av}}$) and the averaged cross section
($\overline{\sigma^{calc}}$) for each bin can be used to correct the
experimental cross section for bin centering
($\sigma^{exp}_{bc}$) 
\begin{equation}
  \sigma^{exp}_{bc} = f_{bc} \cdot \sigma^{exp},\quad
  f_{bc} = \frac{\sigma^{calc}_{kin_{av}}}{\overline{\sigma^{calc}}}\,.
\label{eq:fbc}
\end{equation}
Bin centering corrections (\ref{eq:fbc}) are typically of the order of
a few percent and are larger at the edge of the acceptance compared to
its center (figure~\ref{fig:bin_center_all}). The largest shifts
occur for the $\thetanpcm = 0\degree$ data set. The ratios
calculated using PWIA are considerably smaller and are also shown in
figure~\ref{fig:bin_center_all}.

\begin{figure}
\includegraphics[width=0.5\textwidth,clip=true]{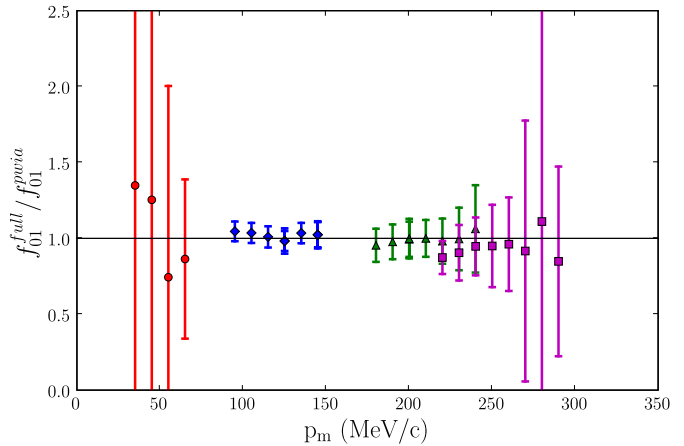}
\caption{\label{fig:mod_dep}Ratio between the experimental values of
  the response function $\flt$ determined using the full 
  calculation for all corrections and using PWIA for the
  corrections. For comparison the error bars indicate the relative
  error in $\flt$ (color online).}
\end{figure}

The bin-centered, 'experimental' cross sections $\sigma^{exp}_{bc}$
(\ref{eq:fbc}) have now well defined kinematic variables that can be
used for comparison with theory without the need to perform a
Monte-Carlo averaging of the theoretical cross sections.  The drawback
of bin center corrections is that one introduces a certain amount of
model-dependence. Comparing the bin-centering corrections between the
full calculation and PWIA gives an estimate of the model
dependence of this approach
(figure~\ref{fig:bin_center_all},\ref{fig:mod_dep}).

When we used the theoretical cross sections for the extraction of
$\flt$ to test the extraction method we found deviations of up to 15~\%
between the obtained value for $\flt$ and the theoretical one. This
was due to the mis-match in the kinematical variables between the
$\phi = 0\degree$ and the $\phi = 180\degree$ kinematic settings for
each corresponding missing momentum bin. These differences affect
especially the photon density matrix $\rho_{ij}$ and the Mott cross
section which should be independent of $\phi$.

The same model used to determine the bin centering correction has
therefore been applied in a second step to correct for these
kinematical differences for each $\ppm$ bin as follows: we calculated
the cross sections for exactly the same kinematics as the $\phi =
0\degree$ data with $\phi$ changed to the appropriate values
of the $\phi = 180\degree$ data leading to
$\sigma^{calc}_{m}$. The experimental cross sections for the $\phi =
180\degree$ data sets have then been corrected for the kinematic
mis-match by multiplying them bin-wise by
$\sigma^{calc}_{m}/\sigma^{calc}_{kin_{av}}$.
\begin{equation}
\sigma^{exp}_{bc,m,\phi=180} =  \sigma^{exp}_{bc,\phi=180}\cdot
\frac{\sigma^{calc}_{m}}{\sigma^{calc}_{kin_{av}}}
\label{eq:sig_corr}
\end{equation}

The matched cross sections have then been used to determine $\flt$ and
the asymmetry $A_{LT}$ defined as
\begin{equation}
A_{LT} = \frac{(\sigma_{\phi = 0\degree} - \sigma_{\phi=180\degree})}
  {(\sigma_{\phi = 0\degree} + \sigma_{\phi=180\degree})}
\label{eq:alt}
\end{equation}
The experimental results for $\flt$ using the procedure described above
are shown in figure~\ref{fig:rlt1} and the one for $A_{LT}$ are shown
in figure~\ref{fig:alt1}.   

To estimate the effect of the model
dependence of the entire procedure on the extraction of $\flt$ we have
performed the same analysis using PWIA in order to calculate the
theoretical cross sections. The ratio between $\flt$ obtained using
the full calculation which includes FSI, MEC, IC and RC and $\flt$
obtained using PWIA is shown in figure~\ref{fig:mod_dep}. In general
the observed deviations are considerably smaller than the error of
$\flt$ and except for the edges of the acceptance smaller than
10~\%. For the comparisons with the calculations we always
used those experimental values that have been extracted using the full
calculation for the correction factors.
\section{Results and Discussion}
\begin{figure}
\includegraphics[width=0.5\textwidth,clip=true]{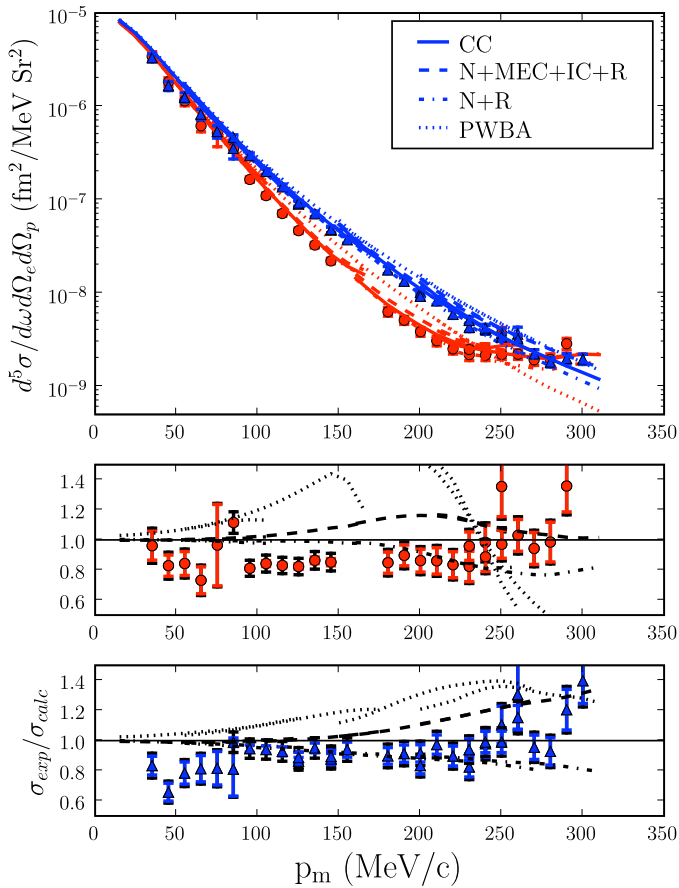}
\caption{\label{fig:sig1}Top: Experimental cross sections as a
  function of missing momentum; red circles correspond to the $\phi
  = 0\degree$ settings and blue triangles correspond to the $\phi =
  180\degree$ kinematics. The lines represent theoretical calculations;
  coupled channel calculation (CC, solid), PWBA (dotted),
  FSI and relativistic corrections (N+R, dash-dot) and FSI, MEC, IC and
  relativistic corrections (N+MEC+IC+R, dashed). Middle: ratio between
  experimental cross sections and calculation for $\phi =
  0\degree$. Bottom: ratio between experimental cross sections and
  calculation for $\phi = 180\degree$. The labeling of the
  calculations is the same as for the top panel (color online).}
\end{figure}

A comparison of the calculated cross sections to the experimental ones
as a function of missing momentum is presented in
figure~\ref{fig:sig1}. The red circles correspond to kinematic
settings centered around $\phi = 0\degree$ and the blue triangles
correspond to kinematic settings centered around $\phi =
180\degree$. In order to better compare the experimental cross
sections to the calculated ones the ratio between experiment and
calculation is shown in the lower two graphs in figure~\ref{fig:sig1}.
While the general behavior of the cross sections as a function of
missing momentum is well reproduced by the calculation, one finds that the
experimental cross sections are generally of the order of 10 to 20~\%
below the calculation especially for the $\phi = 0\degree$
kinematics. This behavior is similar to what has been observed in
other experiments as well and needs further study~\cite{Bern81,bl98, Ulm02}.

The coupled channel calculation, including explicit pionic degrees of
freedom and using the Bonn-OBEPR potential seems to agree best with
the experimental data (solid lines). This is the same calculation that
was compared to experimental results in a determination of the
interference response function $\ftt$ in the delta region by
Pellegrino \etal~\cite{Pel97}. Among the impulse approximation based
calculations the best agreement is obtained by the calculation
including FSI and RC (N+R), while the one including all contributions
(N+MEC+IC+R), systematically over-predicts the cross sections for $\phi
= 0\degree$ as well as for $\phi = 180\degree$.
\begin{figure}
\includegraphics[width=0.5\textwidth,clip=true]{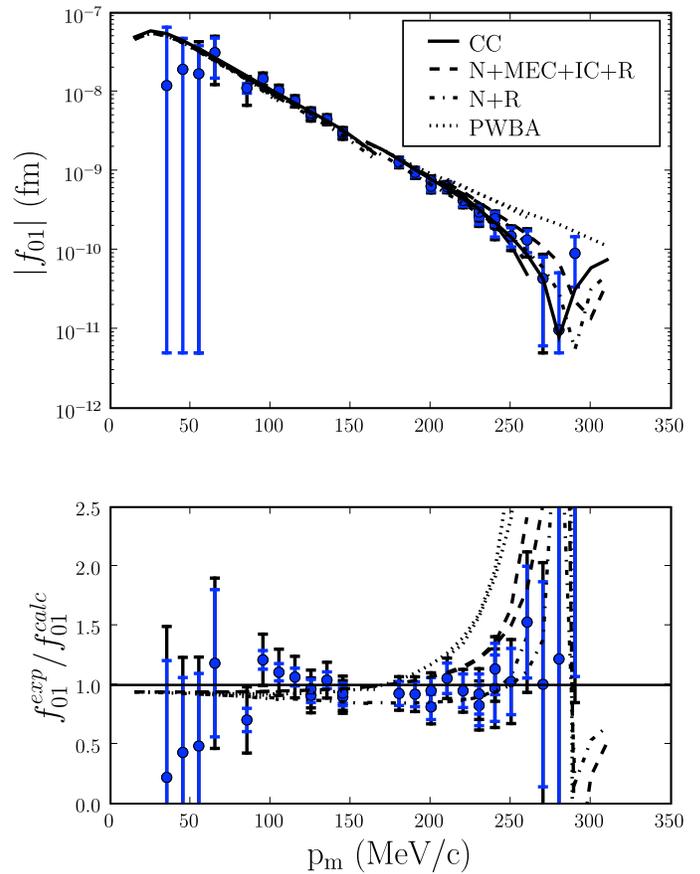}
\caption{\label{fig:rlt1}Top: Extracted interference response function
  $\flt$ as a function of missing momentum. The smaller error bar
  represents the statistical error and the larger one includes the
  systematic uncertainty. As in figure~\ref{fig:sig1} the lines
  represent the theoretical calculations. Bottom: ratio between
  experimental values of $\flt$ and calculations. The labeling of the
  calculations is the same as for the top panel (color online).}
\end{figure}
In figure~\ref{fig:alt1} the extracted asymmetry $A_{LT}$
(\ref{eq:alt}) is compared to the one determined from the coupled
channel model (CC, solid curve). The calculation systematically
deviates from the experiment in the same region where the cross
sections deviate. This indicates that this discrepancy is not due to
an overall normalization factor in the cross sections since an overall
factor would cancel in $A_{LT}$. The observed deviation could be due
to a discrepancy in the interference response function or the
longitudinal (\fl) and/or the transverse (\ft) responses. We compared
the extracted response function $\flt$ to the one calculated from the
coupled channel calculation (figure~\ref{fig:rlt1}) and found that the
calculation agrees well with the experiment within the
experimental error bars. This suggests that the observed differences in the
cross section is not due to a difference in $\flt$ but due to a
discrepancy in the longitudinal/transverse responses.

This is in agreement with the experiments mentioned in the introduction
that extracted $\fl$ and $\ft$ and found that the experimental value of
$\fl$ is systematically smaller than the calculated $\fl$~\cite{Jor96, Duc94}. 
\begin{figure}
\includegraphics[width=0.5\textwidth,clip=true]{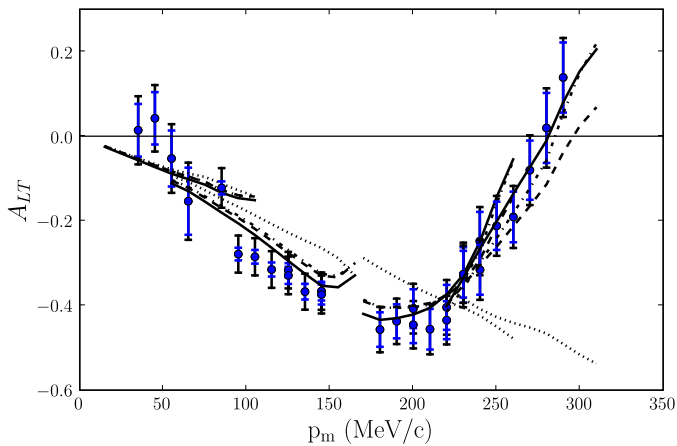}
\caption{\label{fig:alt1} The asymmetry $A_{LT}$ compared to
  the calculation. The labeling of the curves is the same as
  in figures~\ref{fig:sig1},\ref{fig:rlt1}. Both the statistical and the
  total errors are indicated (color online).}
\end{figure}
 
\section{Summary and Conclusion}
In summary, we have measured the $\deep$ cross section for missing
momenta up to 290 MeV/c at $\phi = 0\degree$ and $\phi=180\degree$ and
extracted the interference response function $\flt$. This response is
well reproduced by the coupled channel calculation 
using the Bonn potential. The measured cross sections are
systematically below the calculations by 10 - 20~\% for missing momenta
between 100 MeV/c and 200 MeV/c. This same behavior can also be
observed in other experiments~\cite{Bern81, bl98, Ulm02}. The fact
that the extracted experimental values for $\flt$ agree well with the
calculation leads to the conclusion that the cross section deviations
at lower missing momenta are due to differences in the longitudinal
and/or the transverse responses. To further investigate this issue
would require an L/T separation which has not been carried out for this
kinematics to date.

\begin{acknowledgments}
  Numerical results in electronic form are available from the first
  author.  This work was supported in part by the Department of
  Energy, DOE grant DE--FG02--99ER41065 and by the Deutsche
  Forschungsgemeinschaft (SFB 201)
\end{acknowledgments}
\appendix
\section{Kinematics Tables and Numerical Results}
Tables~\ref{tab:00_ks} -- \ref{tab:18048_ks} contain the average
kinematic setting for each bin in missing momentum together with the
experimental cross section, the bin corrected cross section and the
statistical and systematic errors. Tables~\ref{tab:00_ra} --
\ref{tab:048_ra} contain the extracted response function $\flt$ including
statistical and systematic errors, the asymmetry $A_{LT}$ including
all its errors and the corrected experimental cross section for the $\phi =
180\degree$ setting matched to the kinematics at $\phi = 0\degree$ as
described in the text above.

\begin{table*} 
\caption{\label{tab:00_ks} Averaged kinematics and cross sections for setting : rlt00 } 
\begin{ruledtabular} 
\begin{tabular}{ccccccccccc} 
$\ppm$ & $\theta_e$ & $\omega$ & $q_{lab}$ & 
$\theta_{pq}^{lab}$ & $\phi_{pq}$ & $p_f$ & 
$\sigma_{exp}$ & $\sigma_{exp}^{bin corr.}$ & $\pm \Delta \sigma_{stat}$ & 
$\pm \Delta\sigma_{sys}$ \\ 
{\small $\rm (MeV/c) $} & {\small $(\degree)$} & {\small (MeV)} & {\small $\rm (MeV/c) $} & {\small $(\degree)$} & {\small $(\degree)$} & {\small $\rm (MeV/c) $}& 
{\small $\rm (\frac{fm^2}{(MeV\cdot Sr^2)})$} & {\small $\rm (\frac{fm^2}{(MeV\cdot Sr^2)})$} & {\small $\rm (\frac{fm^2}{(MeV\cdot Sr^2)})$} & {\small $\rm (\frac{fm^2}{(MeV\cdot Sr^2)})$}\\ \hline 
 35.0 & 45.17 & 190.0 & 609.5 &  3.07 & 32.91 & 621.3 &    3.49e-06 &   3.44e-06 &    3.5e-07 &    2.4e-07 \\ 
 45.0 & 45.15 & 193.3 & 609.0 &  3.85 & 32.46 & 626.4 &    1.90e-06 &   1.83e-06 &    1.6e-07 &    1.3e-07 \\ 
 55.0 & 45.07 & 196.3 & 607.9 &  4.62 & 33.55 & 630.9 &    1.22e-06 &   1.15e-06 &    1.2e-07 &    7.9e-08 \\ 
 65.0 & 44.99 & 200.0 & 606.7 &  5.34 & 38.75 & 636.3 &    6.89e-07 &   6.19e-07 &    8.4e-08 &    4.4e-08 \\ 
 75.0 & 44.88 & 201.7 & 605.3 &  6.23 & 45.38 & 637.9 &    5.88e-07 &   5.18e-07 &    1.6e-07 &    3.7e-08 \\ 
\end{tabular} 
\end{ruledtabular} 
\end{table*} 
\begin{table*} 
\caption{\label{tab:020_ks} Averaged kinematics and cross sections for setting : rlt020 } 
\begin{ruledtabular} 
\begin{tabular}{ccccccccccc} 
$\ppm$ & $\theta_e$ & $\omega$ & $q_{lab}$ & 
$\theta_{pq}^{lab}$ & $\phi_{pq}$ & $p_f$ & 
$\sigma_{exp}$ & $\sigma_{exp}^{bin corr.}$ & $\pm \Delta \sigma_{stat}$ & 
$\pm \Delta\sigma_{sys}$ \\ 
{\small $\rm (MeV/c) $} & {\small $(\degree)$} & {\small (MeV)} & {\small $\rm (MeV/c) $} & {\small $(\degree)$} & {\small $(\degree)$} & {\small $\rm (MeV/c) $}& 
{\small $\rm (\frac{fm^2}{(MeV\cdot Sr^2)})$} & {\small $\rm (\frac{fm^2}{(MeV\cdot Sr^2)})$} & {\small $\rm (\frac{fm^2}{(MeV\cdot Sr^2)})$} & {\small $\rm (\frac{fm^2}{(MeV\cdot Sr^2)})$}\\ \hline 
 85.0 & 44.92 & 199.4 & 605.8 &  7.48 &  8.35 & 632.3 &    3.67e-07 &   3.55e-07 &    4.6e-09 &    2.3e-08 \\ 
 95.0 & 44.91 & 202.6 & 605.5 &  8.30 &  8.39 & 636.3 &    1.73e-07 &   1.65e-07 &    3.3e-09 &    1.1e-08 \\ 
105.0 & 44.90 & 204.8 & 605.3 &  9.20 &  8.27 & 638.3 &    1.17e-07 &   1.11e-07 &    2.7e-09 &    7.3e-09 \\ 
115.0 & 44.89 & 204.7 & 605.2 & 10.25 &  8.27 & 636.0 &    7.51e-08 &   7.11e-08 &    1.4e-09 &    4.7e-09 \\ 
125.0 & 44.89 & 204.5 & 605.2 & 11.29 &  8.19 & 633.4 &    4.91e-08 &   4.66e-08 &    1.1e-09 &    3.1e-09 \\ 
135.0 & 44.89 & 204.2 & 605.3 & 12.33 &  8.15 & 630.3 &    3.46e-08 &   3.27e-08 &    8.9e-10 &    2.2e-09 \\ 
145.0 & 44.86 & 200.4 & 605.2 & 13.50 &  8.27 & 621.1 &    2.36e-08 &   2.21e-08 &    6.6e-10 &    1.5e-09 \\ 
\end{tabular} 
\end{ruledtabular} 
\end{table*} 
\begin{table*} 
\caption{\label{tab:040_ks} Averaged kinematics and cross sections for setting : rlt040 } 
\begin{ruledtabular} 
\begin{tabular}{ccccccccccc} 
$\ppm$ & $\theta_e$ & $\omega$ & $q_{lab}$ & 
$\theta_{pq}^{lab}$ & $\phi_{pq}$ & $p_f$ & 
$\sigma_{exp}$ & $\sigma_{exp}^{bin corr.}$ & $\pm \Delta \sigma_{stat}$ & 
$\pm \Delta\sigma_{sys}$ \\ 
{\small $\rm (MeV/c) $} & {\small $(\degree)$} & {\small (MeV)} & {\small $\rm (MeV/c) $} & {\small $(\degree)$} & {\small $(\degree)$} & {\small $\rm (MeV/c) $}& 
{\small $\rm (\frac{fm^2}{(MeV\cdot Sr^2)})$} & {\small $\rm (\frac{fm^2}{(MeV\cdot Sr^2)})$} & {\small $\rm (\frac{fm^2}{(MeV\cdot Sr^2)})$} & {\small $\rm (\frac{fm^2}{(MeV\cdot Sr^2)})$}\\ \hline 
180.0 & 44.89 & 204.9 & 605.1 & 16.88 &  7.52 & 618.2 &    6.48e-09 &   6.30e-09 &    5.4e-10 &    4.0e-10 \\ 
190.0 & 44.89 & 201.1 & 605.5 & 18.02 &  7.52 & 607.7 &    5.40e-09 &   5.12e-09 &    4.2e-10 &    3.4e-10 \\ 
200.0 & 44.90 & 200.4 & 605.6 & 19.05 &  7.52 & 602.6 &    4.11e-09 &   3.85e-09 &    3.6e-10 &    2.6e-10 \\ 
210.0 & 44.90 & 201.2 & 605.6 & 20.05 &  7.39 & 600.1 &    3.29e-09 &   3.08e-09 &    3.3e-10 &    2.0e-10 \\ 
220.0 & 44.90 & 201.1 & 605.6 & 21.08 &  7.25 & 595.9 &    2.68e-09 &   2.52e-09 &    2.8e-10 &    1.7e-10 \\ 
230.0 & 44.90 & 200.8 & 605.6 & 22.12 &  7.11 & 590.8 &    2.31e-09 &   2.19e-09 &    2.8e-10 &    1.4e-10 \\ 
240.0 & 44.90 & 197.9 & 605.8 & 23.21 &  7.16 & 581.1 &    2.60e-09 &   2.52e-09 &    3.0e-10 &    1.6e-10 \\ 
250.0 & 44.92 & 192.5 & 606.4 & 24.31 &  7.16 & 566.3 &    3.69e-09 &   3.69e-09 &    5.4e-10 &    2.3e-10 \\ 
\end{tabular} 
\end{ruledtabular} 
\end{table*} 
\begin{table*} 
\caption{\label{tab:048_ks} Averaged kinematics and cross sections for setting : rlt048 } 
\begin{ruledtabular} 
\begin{tabular}{ccccccccccc} 
$\ppm$ & $\theta_e$ & $\omega$ & $q_{lab}$ & 
$\theta_{pq}^{lab}$ & $\phi_{pq}$ & $p_f$ & 
$\sigma_{exp}$ & $\sigma_{exp}^{bin corr.}$ & $\pm \Delta \sigma_{stat}$ & 
$\pm \Delta\sigma_{sys}$ \\ 
{\small $\rm (MeV/c) $} & {\small $(\degree)$} & {\small (MeV)} & {\small $\rm (MeV/c) $} & {\small $(\degree)$} & {\small $(\degree)$} & {\small $\rm (MeV/c) $}& 
{\small $\rm (\frac{fm^2}{(MeV\cdot Sr^2)})$} & {\small $\rm (\frac{fm^2}{(MeV\cdot Sr^2)})$} & {\small $\rm (\frac{fm^2}{(MeV\cdot Sr^2)})$} & {\small $\rm (\frac{fm^2}{(MeV\cdot Sr^2)})$}\\ \hline 
230.0 & 45.00 & 202.3 & 606.5 & 22.06 &  6.83 & 593.8 &    2.56e-09 &   2.48e-09 &    2.5e-10 &    1.6e-10 \\ 
240.0 & 45.02 & 200.5 & 607.0 & 23.13 &  6.88 & 585.8 &    2.22e-09 &   2.16e-09 &    2.4e-10 &    1.4e-10 \\ 
250.0 & 45.03 & 200.4 & 607.1 & 24.17 &  6.69 & 580.8 &    2.25e-09 &   2.20e-09 &    2.1e-10 &    1.4e-10 \\ 
260.0 & 45.03 & 200.3 & 607.1 & 25.22 &  6.43 & 575.6 &    2.24e-09 &   2.21e-09 &    2.3e-10 &    1.4e-10 \\ 
270.0 & 45.03 & 200.6 & 607.1 & 26.27 &  6.17 & 571.1 &    1.93e-09 &   1.92e-09 &    2.2e-10 &    1.2e-10 \\ 
280.0 & 45.05 & 200.5 & 607.2 & 27.33 &  5.96 & 565.7 &    1.94e-09 &   1.94e-09 &    2.6e-10 &    1.2e-10 \\ 
290.0 & 45.10 & 196.1 & 608.1 & 28.43 &  5.96 & 551.8 &    2.82e-09 &   2.86e-09 &    3.6e-10 &    1.8e-10 \\ 
\end{tabular} 
\end{ruledtabular} 
\end{table*} 
\begin{table*} 
\caption{\label{tab:1800_ks} Averaged kinematics and cross sections for setting : rlt1800 } 
\begin{ruledtabular} 
\begin{tabular}{ccccccccccc} 
$\ppm$ & $\theta_e$ & $\omega$ & $q_{lab}$ & 
$\theta_{pq}^{lab}$ & $\phi_{pq}$ & $p_f$ & 
$\sigma_{exp}$ & $\sigma_{exp}^{bin corr.}$ & $\pm \Delta \sigma_{stat}$ & 
$\pm \Delta\sigma_{sys}$ \\ 
{\small $\rm (MeV/c) $} & {\small $(\degree)$} & {\small (MeV)} & {\small $\rm (MeV/c) $} & {\small $(\degree)$} & {\small $(\degree)$} & {\small $\rm (MeV/c) $}& 
{\small $\rm (\frac{fm^2}{(MeV\cdot Sr^2)})$} & {\small $\rm (\frac{fm^2}{(MeV\cdot Sr^2)})$} & {\small $\rm (\frac{fm^2}{(MeV\cdot Sr^2)})$} & {\small $\rm (\frac{fm^2}{(MeV\cdot Sr^2)})$}\\ \hline 
 35.0 & 45.20 & 191.7 & 609.7 &  2.95 & 145.04 & 624.5 &    3.38e-06 &   3.33e-06 &    2.5e-07 &    2.2e-07 \\ 
 45.0 & 45.15 & 197.7 & 608.5 &  3.39 & 144.54 & 634.4 &    1.75e-06 &   1.66e-06 &    1.6e-07 &    1.1e-07 \\ 
 55.0 & 45.13 & 203.7 & 607.7 &  3.77 & 144.67 & 644.2 &    1.34e-06 &   1.25e-06 &    1.3e-07 &    8.7e-08 \\ 
 65.0 & 45.09 & 209.6 & 606.9 &  4.12 & 145.17 & 653.5 &    8.89e-07 &   8.25e-07 &    9.5e-08 &    5.8e-08 \\ 
 75.0 & 44.99 & 214.0 & 605.5 &  4.69 & 145.29 & 659.8 &    5.88e-07 &   5.38e-07 &    8.0e-08 &    3.8e-08 \\ 
 85.0 & 44.94 & 216.4 & 604.8 &  5.64 & 144.87 & 662.6 &    3.86e-07 &   3.55e-07 &    8.5e-08 &    2.5e-08 \\ 
\end{tabular} 
\end{ruledtabular} 
\end{table*} 
\begin{table*} 
\caption{\label{tab:18020_ks} Averaged kinematics and cross sections for setting : rlt18020 } 
\begin{ruledtabular} 
\begin{tabular}{ccccccccccc} 
$\ppm$ & $\theta_e$ & $\omega$ & $q_{lab}$ & 
$\theta_{pq}^{lab}$ & $\phi_{pq}$ & $p_f$ & 
$\sigma_{exp}$ & $\sigma_{exp}^{bin corr.}$ & $\pm \Delta \sigma_{stat}$ & 
$\pm \Delta\sigma_{sys}$ \\ 
{\small $\rm (MeV/c) $} & {\small $(\degree)$} & {\small (MeV)} & {\small $\rm (MeV/c) $} & {\small $(\degree)$} & {\small $(\degree)$} & {\small $\rm (MeV/c) $}& 
{\small $\rm (\frac{fm^2}{(MeV\cdot Sr^2)})$} & {\small $\rm (\frac{fm^2}{(MeV\cdot Sr^2)})$} & {\small $\rm (\frac{fm^2}{(MeV\cdot Sr^2)})$} & {\small $\rm (\frac{fm^2}{(MeV\cdot Sr^2)})$}\\ \hline 
 85.0 & 44.88 & 191.9 & 606.0 &  7.86 & 171.53 & 619.0 &    4.60e-07 &   4.63e-07 &    1.3e-08 &    2.9e-08 \\ 
 95.0 & 44.89 & 194.9 & 605.8 &  8.73 & 171.61 & 622.5 &    2.96e-07 &   2.96e-07 &    7.7e-09 &    1.9e-08 \\ 
105.0 & 44.90 & 197.9 & 605.7 &  9.60 & 171.65 & 626.0 &    2.02e-07 &   2.00e-07 &    5.3e-09 &    1.3e-08 \\ 
115.0 & 44.90 & 200.5 & 605.6 & 10.48 & 171.69 & 628.6 &    1.38e-07 &   1.37e-07 &    4.5e-09 &    8.7e-09 \\ 
125.0 & 44.90 & 203.2 & 605.4 & 11.36 & 171.69 & 631.0 &    9.09e-08 &   8.96e-08 &    2.7e-09 &    5.7e-09 \\ 
125.0 & 44.90 & 203.2 & 605.4 & 11.36 & 171.69 & 631.0 &    9.09e-08 &   9.22e-08 &    2.7e-09 &    5.7e-09 \\ 
125.0 & 44.90 & 203.2 & 605.4 & 11.36 & 171.69 & 631.0 &    9.35e-08 &   8.96e-08 &    3.7e-09 &    5.9e-09 \\ 
125.0 & 44.90 & 203.2 & 605.4 & 11.36 & 171.69 & 631.0 &    9.35e-08 &   9.22e-08 &    3.7e-09 &    5.9e-09 \\ 
135.0 & 44.91 & 206.4 & 605.3 & 12.21 & 171.65 & 634.4 &    7.16e-08 &   7.05e-08 &    2.4e-09 &    4.5e-09 \\ 
145.0 & 44.92 & 211.1 & 605.0 & 12.97 & 171.65 & 640.3 &    4.75e-08 &   4.72e-08 &    2.0e-09 &    3.0e-09 \\ 
145.0 & 44.92 & 211.1 & 605.0 & 12.97 & 171.65 & 640.3 &    4.75e-08 &   4.81e-08 &    2.0e-09 &    3.0e-09 \\ 
145.0 & 44.92 & 211.1 & 605.0 & 12.97 & 171.65 & 640.3 &    4.84e-08 &   4.72e-08 &    2.3e-09 &    3.0e-09 \\ 
145.0 & 44.92 & 211.1 & 605.0 & 12.97 & 171.65 & 640.3 &    4.84e-08 &   4.81e-08 &    2.3e-09 &    3.0e-09 \\ 
155.0 & 44.92 & 215.2 & 604.7 & 13.77 & 171.69 & 644.9 &    3.71e-08 &   3.72e-08 &    2.0e-09 &    2.3e-09 \\ 
\end{tabular} 
\end{ruledtabular} 
\end{table*} 
\begin{table*} 
\caption{\label{tab:18040_ks} Averaged kinematics and cross sections for setting : rlt18040 } 
\begin{ruledtabular} 
\begin{tabular}{ccccccccccc} 
$\ppm$ & $\theta_e$ & $\omega$ & $q_{lab}$ & 
$\theta_{pq}^{lab}$ & $\phi_{pq}$ & $p_f$ & 
$\sigma_{exp}$ & $\sigma_{exp}^{bin corr.}$ & $\pm \Delta \sigma_{stat}$ & 
$\pm \Delta\sigma_{sys}$ \\ 
{\small $\rm (MeV/c) $} & {\small $(\degree)$} & {\small (MeV)} & {\small $\rm (MeV/c) $} & {\small $(\degree)$} & {\small $(\degree)$} & {\small $\rm (MeV/c) $}& 
{\small $\rm (\frac{fm^2}{(MeV\cdot Sr^2)})$} & {\small $\rm (\frac{fm^2}{(MeV\cdot Sr^2)})$} & {\small $\rm (\frac{fm^2}{(MeV\cdot Sr^2)})$} & {\small $\rm (\frac{fm^2}{(MeV\cdot Sr^2)})$}\\ \hline 
180.0 & 44.87 & 191.0 & 606.0 & 17.22 & 172.10 & 592.9 &    1.75e-08 &   1.77e-08 &    1.1e-09 &    1.1e-09 \\ 
190.0 & 44.89 & 194.5 & 605.9 & 18.17 & 172.14 & 595.6 &    1.34e-08 &   1.34e-08 &    8.4e-10 &    8.3e-10 \\ 
200.0 & 44.89 & 197.7 & 605.7 & 19.12 & 172.10 & 597.8 &    1.02e-08 &   1.01e-08 &    6.4e-10 &    6.4e-10 \\ 
200.0 & 44.89 & 197.7 & 605.7 & 19.12 & 172.10 & 597.8 &    1.02e-08 &   9.23e-09 &    6.4e-10 &    6.4e-10 \\ 
200.0 & 44.89 & 197.7 & 605.7 & 19.12 & 172.10 & 597.8 &    9.33e-09 &   1.01e-08 &    6.1e-10 &    5.8e-10 \\ 
200.0 & 44.89 & 197.7 & 605.7 & 19.12 & 172.10 & 597.8 &    9.33e-09 &   9.23e-09 &    6.1e-10 &    5.8e-10 \\ 
210.0 & 44.90 & 200.9 & 605.5 & 20.06 & 172.14 & 599.6 &    8.43e-09 &   8.25e-09 &    5.6e-10 &    5.2e-10 \\ 
220.0 & 44.90 & 202.8 & 605.4 & 21.04 & 172.14 & 599.1 &    6.10e-09 &   5.93e-09 &    4.5e-10 &    3.8e-10 \\ 
230.0 & 44.91 & 203.6 & 605.5 & 22.05 & 172.14 & 596.2 &    4.44e-09 &   4.29e-09 &    3.6e-10 &    2.8e-10 \\ 
240.0 & 44.93 & 206.8 & 605.5 & 23.00 & 172.14 & 597.7 &    4.15e-09 &   4.06e-09 &    3.7e-10 &    2.6e-10 \\ 
250.0 & 44.94 & 212.1 & 605.2 & 23.88 & 172.14 & 603.2 &    3.70e-09 &   3.71e-09 &    4.2e-10 &    2.3e-10 \\ 
260.0 & 44.97 & 219.4 & 605.1 & 24.65 & 172.31 & 612.3 &    3.59e-09 &   3.63e-09 &    6.3e-10 &    2.2e-10 \\ 
\end{tabular} 
\end{ruledtabular} 
\end{table*} 
\begin{table*} 
\caption{\label{tab:18048_ks} Averaged kinematics and cross sections for setting : rlt18048 } 
\begin{ruledtabular} 
\begin{tabular}{ccccccccccc} 
$\ppm$ & $\theta_e$ & $\omega$ & $q_{lab}$ & 
$\theta_{pq}^{lab}$ & $\phi_{pq}$ & $p_f$ & 
$\sigma_{exp}$ & $\sigma_{exp}^{bin corr.}$ & $\pm \Delta \sigma_{stat}$ & 
$\pm \Delta\sigma_{sys}$ \\ 
{\small $\rm (MeV/c) $} & {\small $(\degree)$} & {\small (MeV)} & {\small $\rm (MeV/c) $} & {\small $(\degree)$} & {\small $(\degree)$} & {\small $\rm (MeV/c) $}& 
{\small $\rm (\frac{fm^2}{(MeV\cdot Sr^2)})$} & {\small $\rm (\frac{fm^2}{(MeV\cdot Sr^2)})$} & {\small $\rm (\frac{fm^2}{(MeV\cdot Sr^2)})$} & {\small $\rm (\frac{fm^2}{(MeV\cdot Sr^2)})$}\\ \hline 
230.0 & 45.00 & 194.3 & 607.2 & 22.20 & 172.39 & 578.8 &    5.13e-09 &   5.10e-09 &    3.8e-10 &    3.2e-10 \\ 
240.0 & 45.00 & 198.1 & 606.9 & 23.18 & 172.39 & 581.4 &    4.28e-09 &   4.21e-09 &    3.1e-10 &    2.7e-10 \\ 
250.0 & 45.01 & 201.1 & 606.7 & 24.16 & 172.39 & 582.2 &    3.46e-09 &   3.37e-09 &    2.5e-10 &    2.2e-10 \\ 
260.0 & 45.02 & 202.1 & 606.8 & 25.19 & 172.39 & 579.2 &    3.33e-09 &   3.23e-09 &    2.3e-10 &    2.1e-10 \\ 
270.0 & 45.03 & 202.5 & 606.9 & 26.24 & 172.44 & 574.6 &    2.31e-09 &   2.23e-09 &    2.0e-10 &    1.4e-10 \\ 
280.0 & 45.06 & 204.5 & 607.0 & 27.25 & 172.48 & 573.3 &    1.87e-09 &   1.82e-09 &    1.8e-10 &    1.2e-10 \\ 
290.0 & 45.09 & 208.8 & 607.0 & 28.22 & 172.48 & 576.2 &    2.01e-09 &   1.99e-09 &    2.3e-10 &    1.3e-10 \\ 
300.0 & 45.10 & 214.1 & 606.8 & 29.16 & 172.44 & 580.8 &    1.95e-09 &   1.96e-09 &    2.1e-10 &    1.2e-10 \\ 
\end{tabular} 
\end{ruledtabular} 
\end{table*} 

\begin{table*} 
\caption{\label{tab:00_ra} $\flt$ and $A_{LT}$ from settings : rlt00 and rlt1800 } 
\begin{ruledtabular} 
\begin{tabular}{cccccccc} 
$\ppm$ & $\flt $ & $\Delta \flt_{stat}$ & $\Delta \flt_{sys}$ & 
$A_{LT}$ & $\Delta A_{LT, stat}$ & $\Delta A_{LT, sys} $ & $\sigma_{exp, m}$ \\\ 
{\small $\rm (MeV/c) $} & {\small $\rm (fm)$} & {\small $\rm (fm)$} & {\small $\rm (fm)$} &    &    &   & {\small $\rm (\frac{fm^2}{(MeV\cdot Sr^2)})$} \\ \hline 
  35.0 &   1.20e-08 &    5.4e-08 &    4.4e-08 &   1.40e-02  &    6.2e-02 &    5.1e-02 &   3.34e-06 \\ 
  45.1 &   1.92e-08 &    2.8e-08 &    2.2e-08 &   4.23e-02  &    6.2e-02 &    4.8e-02 &   1.68e-06 \\ 
  55.0 &  -1.69e-08 &    2.1e-08 &    1.5e-08 &  -5.27e-02  &    6.6e-02 &    4.7e-02 &   1.27e-06 \\ 
  65.0 &  -3.14e-08 &    1.6e-08 &    9.6e-09 &  -1.54e-01  &    7.9e-02 &    4.5e-02 &   8.43e-07 \\ 
\end{tabular} 
\end{ruledtabular} 
\end{table*} 
\begin{table*} 
\caption{\label{tab:020_ra} $\flt$ and $A_{LT}$ from settings : rlt020 and rlt18020 } 
\begin{ruledtabular} 
\begin{tabular}{cccccccc} 
$\ppm$ & $\flt $ & $\Delta \flt_{stat}$ & $\Delta \flt_{sys}$ & 
$A_{LT}$ & $\Delta A_{LT, stat}$ & $\Delta A_{LT, sys} $ & $\sigma_{exp, m}$ \\\ 
{\small $\rm (MeV/c) $} & {\small $\rm (fm)$} & {\small $\rm (fm)$} & {\small $\rm (fm)$} &    &    &   & {\small $\rm (\frac{fm^2}{(MeV\cdot Sr^2)})$} \\ \hline 
  85.0 &  -1.12e-08 &    1.6e-09 &    4.1e-09 &  -1.22e-01  &    1.5e-02 &    4.4e-02 &   4.54e-07 \\ 
  95.1 &  -1.47e-08 &    9.5e-10 &    2.4e-09 &  -2.79e-01  &    1.5e-02 &    4.1e-02 &   2.92e-07 \\ 
 105.1 &  -1.03e-08 &    6.8e-10 &    1.7e-09 &  -2.85e-01  &    1.6e-02 &    4.1e-02 &   1.99e-07 \\ 
 115.1 &  -7.68e-09 &    5.4e-10 &    1.1e-09 &  -3.15e-01  &    1.7e-02 &    4.0e-02 &   1.36e-07 \\ 
 125.1 &  -5.09e-09 &    3.3e-10 &    7.5e-10 &  -3.16e-01  &    1.6e-02 &    4.0e-02 &   8.97e-08 \\ 
 125.1 &  -5.09e-09 &    3.3e-10 &    7.5e-10 &  -3.16e-01  &    1.6e-02 &    4.0e-02 &   9.22e-08 \\ 
 125.1 &  -5.39e-09 &    4.5e-10 &    7.7e-10 &  -3.29e-01  &    2.0e-02 &    4.0e-02 &   8.97e-08 \\ 
 125.1 &  -5.39e-09 &    4.5e-10 &    7.7e-10 &  -3.29e-01  &    2.0e-02 &    4.0e-02 &   9.22e-08 \\ 
 135.1 &  -4.49e-09 &    3.0e-10 &    5.8e-10 &  -3.67e-01  &    1.8e-02 &    3.8e-02 &   7.06e-08 \\ 
 145.1 &  -2.96e-09 &    2.5e-10 &    3.8e-10 &  -3.66e-01  &    2.2e-02 &    3.8e-02 &   4.76e-08 \\ 
 145.1 &  -2.96e-09 &    2.5e-10 &    3.8e-10 &  -3.66e-01  &    2.2e-02 &    3.8e-02 &   4.85e-08 \\ 
 145.1 &  -3.07e-09 &    2.7e-10 &    3.9e-10 &  -3.74e-01  &    2.4e-02 &    3.8e-02 &   4.76e-08 \\ 
 145.1 &  -3.07e-09 &    2.7e-10 &    3.9e-10 &  -3.74e-01  &    2.4e-02 &    3.8e-02 &   4.85e-08 \\ 
\end{tabular} 
\end{ruledtabular} 
\end{table*} 
\begin{table*} 
\caption{\label{tab:040_ra} $\flt$ and $A_{LT}$ from settings : rlt040 and rlt18040 } 
\begin{ruledtabular} 
\begin{tabular}{cccccccc} 
$\ppm$ & $\flt $ & $\Delta \flt_{stat}$ & $\Delta \flt_{sys}$ & 
$A_{LT}$ & $\Delta A_{LT, stat}$ & $\Delta A_{LT, sys} $ & $\sigma_{exp, m}$ \\\ 
{\small $\rm (MeV/c) $} & {\small $\rm (fm)$} & {\small $\rm (fm)$} & {\small $\rm (fm)$} &    &    &   & {\small $\rm (\frac{fm^2}{(MeV\cdot Sr^2)})$} \\ \hline 
 180.0 &  -1.29e-09 &    1.4e-10 &    1.4e-10 &  -4.57e-01  &    4.1e-02 &    3.5e-02 &   1.69e-08 \\ 
 190.0 &  -9.57e-10 &    1.1e-10 &    1.1e-10 &  -4.37e-01  &    4.0e-02 &    3.6e-02 &   1.31e-08 \\ 
 200.0 &  -7.47e-10 &    8.7e-11 &    8.1e-11 &  -4.45e-01  &    4.3e-02 &    3.5e-02 &   1.00e-08 \\ 
 200.0 &  -7.47e-10 &    8.7e-11 &    8.1e-11 &  -4.45e-01  &    4.3e-02 &    3.5e-02 &   9.15e-09 \\ 
 200.0 &  -6.39e-10 &    8.3e-11 &    7.5e-11 &  -4.07e-01  &    4.6e-02 &    3.7e-02 &   1.00e-08 \\ 
 200.0 &  -6.39e-10 &    8.3e-11 &    7.5e-11 &  -4.07e-01  &    4.6e-02 &    3.7e-02 &   9.15e-09 \\ 
 210.1 &  -6.31e-10 &    7.7e-11 &    6.7e-11 &  -4.56e-01  &    4.8e-02 &    3.5e-02 &   8.25e-09 \\ 
 220.1 &  -4.22e-10 &    6.3e-11 &    5.0e-11 &  -4.04e-01  &    5.3e-02 &    3.7e-02 &   5.95e-09 \\ 
 230.1 &  -2.65e-10 &    5.5e-11 &    3.7e-11 &  -3.28e-01  &    6.5e-02 &    3.9e-02 &   4.33e-09 \\ 
 240.1 &  -2.05e-10 &    5.9e-11 &    3.7e-11 &  -2.48e-01  &    6.9e-02 &    4.1e-02 &   4.19e-09 \\ 
\end{tabular} 
\end{ruledtabular} 
\end{table*} 
\begin{table*} 
\caption{\label{tab:048_ra} $\flt$ and $A_{LT}$ from settings : rlt048 and rlt18048 } 
\begin{ruledtabular} 
\begin{tabular}{cccccccc} 
$\ppm$ & $\flt $ & $\Delta \flt_{stat}$ & $\Delta \flt_{sys}$ & 
$A_{LT}$ & $\Delta A_{LT, stat}$ & $\Delta A_{LT, sys} $ & $\sigma_{exp, m}$ \\\ 
{\small $\rm (MeV/c) $} & {\small $\rm (fm)$} & {\small $\rm (fm)$} & {\small $\rm (fm)$} &    &    &   & {\small $\rm (\frac{fm^2}{(MeV\cdot Sr^2)})$} \\ \hline 
 230.0 &  -3.02e-10 &    5.5e-11 &    4.3e-11 &  -3.26e-01  &    5.5e-02 &    3.9e-02 &   4.88e-09 \\ 
 240.0 &  -2.51e-10 &    4.8e-11 &    3.7e-11 &  -3.16e-01  &    5.9e-02 &    4.0e-02 &   4.15e-09 \\ 
 250.0 &  -1.50e-10 &    4.1e-11 &    3.2e-11 &  -2.12e-01  &    5.7e-02 &    4.2e-02 &   3.38e-09 \\ 
 260.0 &  -1.34e-10 &    4.1e-11 &    3.1e-11 &  -1.91e-01  &    6.0e-02 &    4.2e-02 &   3.25e-09 \\ 
 270.1 &  -4.37e-11 &    3.8e-11 &    2.4e-11 &  -8.03e-02  &    7.0e-02 &    4.4e-02 &   2.26e-09 \\ 
 280.0 &   9.76e-12 &    4.2e-11 &    2.2e-11 &   1.95e-02  &    8.3e-02 &    4.4e-02 &   1.87e-09 \\ 
 290.0 &   9.07e-11 &    5.7e-11 &    2.9e-11 &   1.39e-01  &    8.3e-02 &    4.3e-02 &   2.17e-09 \\ 
\end{tabular} 
\end{ruledtabular} 
\end{table*} 

\clearpage
%
%
\bibliography{all}
\end{document}